\documentclass[conference]{IEEEtran}
\IEEEoverridecommandlockouts
\usepackage{cite}
\usepackage{amsmath,amssymb,amsfonts}
\usepackage{algorithmic}
\usepackage{graphicx}
\usepackage{textcomp}
\usepackage{xcolor}
\usepackage{cite}
\usepackage{algorithmic}
\usepackage{algorithm}
\usepackage{enumerate}
\usepackage{subcaption}
\def\BibTeX{{\rm B\kern-.05em{\sc i\kern-.025em b}\kern-.08em
    T\kern-.1667em\lower.7ex\hbox{E}\kern-.125emX}}
\begin{document}

\title{Constrained Deployment Optimization in Integrated Access and Backhaul Networks\\
{\footnotesize \textsuperscript{}}
%\thanks{Identify applicable funding agency here. If none, delete this.}
}

\author{\IEEEauthorblockN{1\textsuperscript{st} Charitha Madapatha}
\IEEEauthorblockA{\textit{Dept. of Electrical Engineering} \\
\textit{Chalmers University of Technology}\\
Gothenburg, Sweden \\
charitha@chalmers.se}
\and
\IEEEauthorblockN{2\textsuperscript{nd} Behrooz Makki}
\IEEEauthorblockA{\textit{Ericsson Research} \\
\textit{Ericsson AB}\\
Gothenburg, Sweden \\
behrooz.makki@ericsson.com}
\and
\IEEEauthorblockN{3\textsuperscript{rd} Hao Guo}
\IEEEauthorblockA{\textit{Dept. of Electrical Engineering} \\
\textit{Chalmers University of Technology}\\
Gothenburg, Sweden \\
hao.guo@chalmers.se}
\and
\IEEEauthorblockN{\centerline{4\textsuperscript{th} Tommy Svensson}
\IEEEauthorblockA{\textit{Dept. of Electrical Engineering} \\
\textit{Chalmers University of Technology}\\
Gothenburg, Sweden \\
tommy.svensson@chalmers.se}}

}

\maketitle

\begin{abstract}
Integrated access and
backhaul (IAB) is one of the promising techniques for 5G networks and beyond (6G), in which the same node/hardware is used to provide both backhaul and cellular services in a multi-hop fashion. Due to the sensitivity of the backhaul links with high rate/reliability demands, proper network planning is needed to make the IAB network performing appropriately and as good as possible. In this paper, we study the effect of deployment optimization on the coverage of IAB networks. We concentrate on the cases where,  due to either geographical or interference management limitations, unconstrained IAB node placement is not feasible in some areas. To that end, we propose various millimeter wave (mmWave) blocking-aware constrained deployment optimization approaches. Our results indicate that, even with limitations on deployment optimization, network planning boosts the coverage of IAB networks considerably.
\end{abstract}

\begin{IEEEkeywords}
Integrated access and backhaul, IAB, Topology optimization, \textcolor{black}{Densification},  millimeter wave (mmWave) communications, 3GPP, Coverage, Wireless backhaul, 5G NR, 6G, Blockage, Machine learning, Network planning.
\end{IEEEkeywords}

\section{Introduction}
The data traffic and the users' rate/reliability demands continue to steadily increase in 5G and beyond  (6G) \cite{rajatheva2020scoring}. In order to meet such demands, network densification, i.e, the deployment of many base stations (BSs) of different types is one of the key enablers. These increasing number of BSs, however, need to be connected to the core network using the transport network.

According to \cite{eref2}, the backhaul technology varies across different regions. However, optical fiber and microwave links have been globally the dominating media for the backhaul. Recently, fiber deployments have increased due to their reliability, and have demonstrated Tbps-level data rates. On the other hand, due to low initial investment and installation time, wireless backhaul comes with considerably lower price, flexibility and time-to-market, at the cost of low peak rate.

Typical wireless backhaul technologies are mainly based on 1) point-to-point line-of-sight (LoS) communications in the range of 10-80 GHz, 2) non-standardized solutions, and 3) accurate network planning such that the interference to/from the backhaul transceivers is minimized. With 5G, however, access communication, i.e., the communication between the gNB and the user equipments (UEs), moves to the millimeter wave (mmWave) band, i.e., the band which was previously used for backhauling. Thus, there will be conflict of interest between access and backhaul, which requires coordination. Also, considering small access points on, e.g., lamppost, one needs to support NLoS (N: non) communication in (possibly, unplanned) backhaul networks. These are the main motivations for the so called integrated access and backhaul (IAB) where the operators can use portion of the radio network resources for wireless backhaul. That is, IAB provides not only access link cellular service but also backhaul using the same node. IAB has been standardized for 5G NR in 3GPP Release-16, Release-17 \cite{3gppnew2}, \cite{3gppnewc2} and, the standardization will be continued in Release-18 %\cite{3gppnew1,madapatha2020integrated,monteiro2022tdd,3gppnew2,3gppnew3}. 
\cite{monteiro2022tdd,3gppnewc1,monteiro2022paving}.

IAB network supports multi-hop communication in which an IAB donor, connected to the core network via, e.g., a fiber link, includes a central unit (CU) for the following concatenated IAB nodes which are connected to IAB donor in a multi-hop fashion (see Fig. 1). Each IAB node consists of two modules, namely, mobile termination (MT) and distributed unit (DU). The DU part of an IAB node is used to serve UEs or the MT part of \textit{child} IAB nodes. The MT part of the IAB is used to connect the IAB node to its \textit{parent} IAB-DU in the multi-hop chain towards the IAB donor. In general, the DU part has similar gNB functionalities, although there may be IAB-specific differences. The IAB-MT part, on the other hand, may have different capabilities, although in general it acts not differently from a UE from the point-of-view of its parent IAB.

In practice, IAB networks may face deployment constraints, where the nodes can not be deployed in some locations. Such constraints may come from two reasons: On one hand, depending on the location and regulatory restrictions in protected areas, it may not be possible/allowed to have the IAB nodes in, e.g., some areas. Although these restrictions vary based on the country and locality, all provinces have their own building and
landscape protection laws. Additionally, federal laws have to be obeyed and permissions under these laws, if applicable,
have to be obtained (e.g. air traffic safety, forest protection,
listed buildings etc.). On the other hand, network planning may impose constraints on IAB nodes placement, e.g., to limit the interference. For instance, 3GPP has defined two categories of IAB nodes, namely, wide- and local-area IAB, with distinct properties \cite{3gppnew1,ronkainen2021integrated}. The main differences between these two categories are in the nodes capabilities and the level of required network planning.

\textcolor{black}{
Wide-area IAB-node can be seen as an independent IAB-node providing its own coverage, with possibly long backhaul link to connect to its parent IAB-node. Here, the goal is to extend the coverage. Due to radio frequency properties, wide-area IAB-node deployment are \textit{well-planned}, by operators. For these type of IAB-nodes, the MT part of the IAB node looks like a normal gNB, in terms of, e.g., high transmit power, beamforming or antenna gains. In wide-area IAB networks, one may consider a minimum distance between the nodes with, e.g., LOS connections. On the other hand, the use-case for the local-area IAB-node is to boost the capacity within an already existing cell served by an IAB donor or parent IAB-node. With local-area IAB networks, the transmit power of the MT part may range between those of UEs and gNBs. Also, the network may be fairly unplanned, while geographical-based constraints may still prevent unconstrained IAB installation in different places.}

In this paper, we study the effect of network planning on the service coverage of IAB networks. We present different algorithms for constrained deployment optimization, with the constraints coming from either inter-IAB distance limitations or geographical restrictions. Moreover, we study the effect of different parameters on the network performance. As we show, even with constraints on deployment optimization, the coverage of IAB networks can be considerably improved via proper network planning.

Note that the problem of topology optimization in different IAB or non-IAB networks have been previously studied in, e.g., \cite{gen4,gen5,9548327,9437349_2021,lai2020resource,fu2019dynamic}. However, compared to the literature, we present different algorithms for deployment optimization, consider different types of constraints and study the performance of IAB networks with various parameter settings, which makes our paper different from the previous works.

\section{System Model}\label{system_model}
\begin{figure}
\centerline{\includegraphics[width=3.5in]{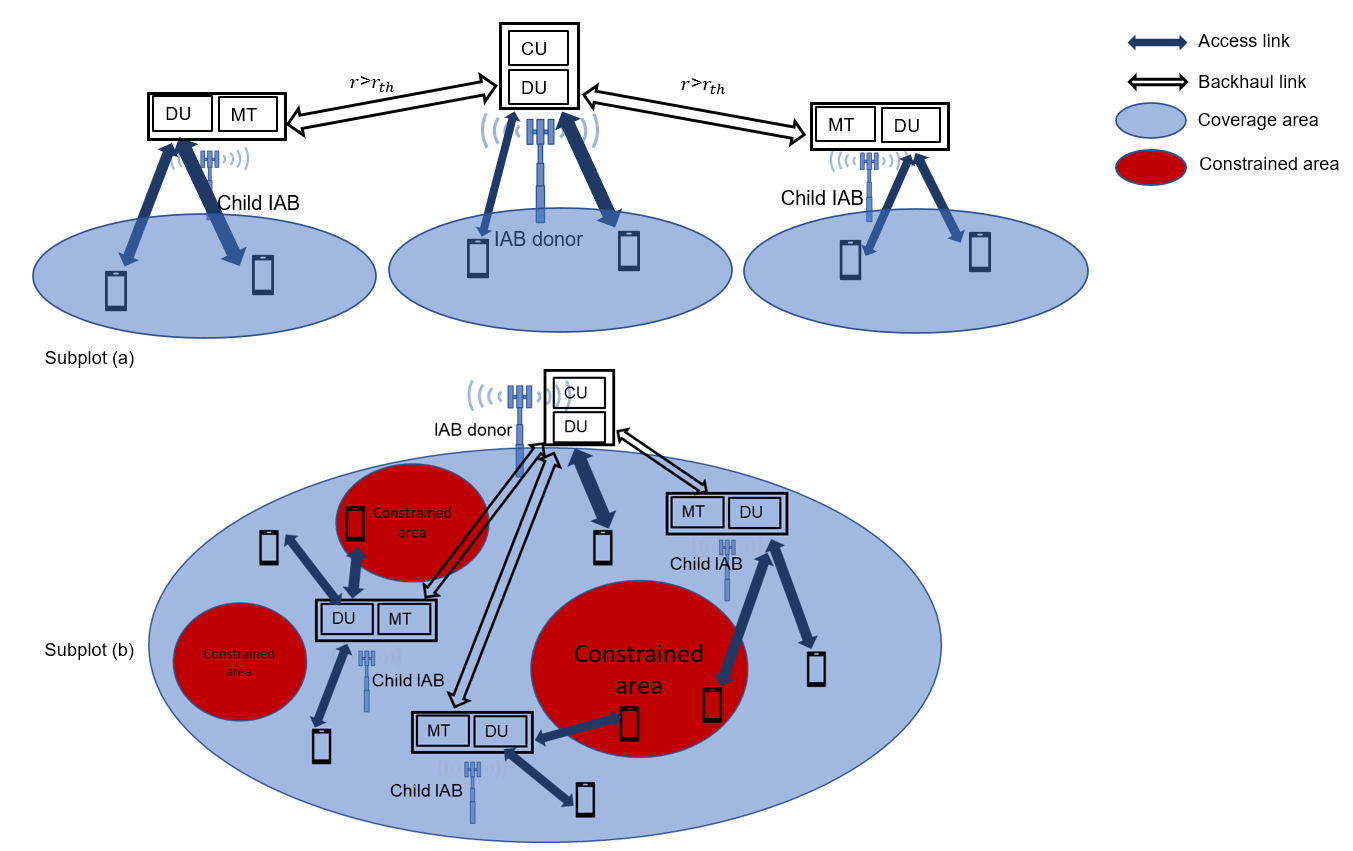}}
\caption{\textcolor{black}{An illustration of the IAB netowrk. Subplot (a): An IAB network with a minimum required distance between the IAB nodes and the IAB-MTs having gNB-like capabilities. Subplot (b): An IAB network with geographical constraints on node placement and the IAB-MT being less capable compared to an gNB.}}
\label{systemmodel}
\end{figure}

Consider downlink communication in a two-hop IAB network, where the IAB donor and its child IAB nodes serve multiple UEs \cite{singh2015tractable,saha2019millimeter,chen2019intelligent,adare2021uplink,fang2021joint} (see Fig. \ref{systemmodel}). Since in-band communication offers proper flexibility for resource allocation, at the cost of co-ordination complexity, we consider an in-band setup where both access and backhaul links operate over the same mmWave band. 

In one scenario as shown in Fig. 1a, the IAB nodes with gNB-like IAB-MT capabilities maintain a minimum distance $r_\text{th}$ between each other, i.e., the distance between every two node $s$ should be $s>r_\text{th}$ where $r_\text{th}$ is a threshold distance considered by the network designer, when there is no blockage in the links between IAB nodes. In another scenario shown in Fig. 1b, while the IAB nodes can be in different distances to each other, due to geographical or regulatory restrictions, it may not be possible to have the nodes in some specific areas.

We use the germ grain model \cite[Chapter 14]{ref5} to model the blockings which provides accurate blind spot prediction. Particularly, a finite homogeneous poisson point process (FHPPP) is used to model the blockings in an area with the blocking density $\lambda_\text{bl}$. The blockings are considered to be walls of length $l_{\text{bl}}$ and orientation $\theta_{\text{bl}}$.

Using the state-of-the-art mmWave channel model, e.g., \cite{ref1, madapatha2020integrated}, the received power at each node can be described as

\begin{equation}
    P_{\text{r}}=P_{\text{t}}h_{\text{t,r}}L_{\text{tr}}G_{\text{t,r}}\left|\left|x_{\text{t}}-x_{\text{r}}\right|\right|^{-1}.
    \label{eqa1}
\end{equation}
Here, $P_{\text{t}}$ stands for the transmit power, $h_{\text{t,r}}$ denotes the small-scale fading of the link, $L_{\text{t,r}}$ is the path loss  according to 5GCM UMa close-in model described in \cite{ref3}, and $G_{\text{t,r}}$ denotes  the  combined antenna  gain  of  the  transmitter  and  receiver  in  the  link.
Particulary, the antenna gain is characterized according to sectored-pattern  antenna  array  model by
\begin{equation}G_{\text{t,r} }(\alpha) = \begin{cases} G_{\text{m}}&\frac{-\alpha_{\text{HP}}}2\leq\alpha\leq\frac{\alpha_{\text{HP}}}2 \\ G_{\text{s}}&\text{otherwise,} \end{cases}
\end{equation}
where $G_{\text{m}}$ denotes the main lobe antenna gain and $G_{\text{s}}$ represents the side lobe antenna gain. Furthermore, in our two-hop setup, each of the UEs can be connected to either the IAB donor or a child IAB, depending on the received power at the UE. Thereby, the interference observed by UE $u$, caused by the neighbouring interferers, is expressed as    
\begin{equation} I_{{\rm {\textit{u}}}}= \sum \limits _{ {\textit{$\mathbf{i}$}\in \chi _{i,u}\setminus \{{\mathbf{w}}_{u}\}}}{P_{i}}  h_{i,u} G_{i,u} L_{\text{t,r}}\|{\mathbf{x_\textit{i}-x_\textit{u}}}\|^{-1},
\label{inter}
\end{equation}
where $i$ represents the nodes excluding the associated node $w_\text{u}$ of UE $u$. Moreover, for child IAB node $c$, the aggregated interference on the backhaul links is given by
\begin{align} I_{{\rm {\textit{c}}}}= \sum \limits _{ {\textit{$\mathbf{j}$}\in \chi_{j,c}\setminus \{{\mathbf{w}}_{c}\}}}{P_{j}}  h_{j,c}L_{j,c} G_{j,c}\|{\mathbf{x_\textit{j}-x_\textit{c}}}\|^{-1},
\label{intersbs}
\end{align}
where $j$ represents transmitting nodes with the exclusion of associated node $w_\text{c}$ of child node $c$. The available mmWave spectrum is partitioned into access and backhaul links such that

\begin{equation}\begin{aligned}
    \begin{cases}
    W_\text{Backhaul}=\beta W\\
    W_\text{Access}=(1-\beta)W,
\end{cases}
\end{aligned}
  \label{eq:11}  
\end{equation}
where $W$ denotes the bandwidth and $\beta$ $\in [0,1]$ represents the bandwidth partitioning factor. 
\textcolor{black}{With our implementation, the network may have two types of access links, i.e., IAB donor-UE or child IAB-UE, and the individual UE data rate depends on the type of the access link. In particular, the UE data rates in access links that are connected to the IAB donor or to the child IAB nodes are given by }
\begin{equation}
 {R_{{u}}} =   \begin{cases}\frac{(1-\beta)W}{{}N_d}\log(1+\text{SINR}(x_{u})), ~\text { if }{\mathbf{w}}_\text{u}\in \chi _{\mathrm{ d}},\\ \min \bigg (\frac{(1-\beta)WN}{{\displaystyle\sum_{\forall\;u}}N_{j,u}}\log(1+\text{SINR}(x_{u})), \\ \qquad  \frac{\beta WN}{{\displaystyle\sum_{\forall\;j}}N_j}\log(1+\text{SINR}(x_\text{b}))\bigg ), \text {if }{\mathbf{w}}_\text{u}\in \chi _{\mathrm{ c}}, \end{cases}
 \label{eq:14}
\end{equation}
where $j$ denotes each child IAB node connected to the IAB donor $d$, which shares some of its bandwidth with child IAB nodes. Moreover, $c$ denotes the child node, and  $u$ identifies the UE. Thereby, 
$\chi_{\mathrm{d}}$, $\chi_{\mathrm{c}}$, $\chi_{\mathrm{u}}$ denote the set of IAB donors, child IAB nodes and UEs, respectively. In particular, using the rates \eqref{eq:14}, our goal is to perform constrained deployment optimization such that service coverage given by
\begin{align}
 \text{CP}= \Pr(R_\text{U} \ge \rho),
 \label{eq:rho}
\end{align}
is maximized. Here, $\rho$ denotes a minimum rate threshold requirement considered by the network designer.

% \begin{table*}[t]
% \centering
% \caption{The Definition of the Parameters.}
% \label{table}
% \setlength{\tabcolsep}{3pt}
% \begin{tabular}{|p{40pt}|p{160pt}|p{40pt}|p{160pt}|}
% \hline
% Parameter& 
% Definition&
% Parameter&
% Definition\\
% \hline

%  $\lambda_{\text{bl}}$& Blocking density&$\lambda_{\text{T}}$& Tree density \\
%  $\theta$& Orientation of the blocking wall&$\alpha$&Angle between transmitter and receiver\\
%  $P_{\text{t}}$&Transmission power&$P_{\text{r}}$&Received power\\
%  $h$&Fading coeficient&$G$&Antenna gain\\

%  $x$&Location of the node&$r$&Propagation distance  between  the  nodes\\
 
% \hline
% \end{tabular}
% \label{tab1x}
% \end{table*}

\begin{algorithm}[tbph]
\caption{IAB placement with minimum inter-IAB distance requirement}
%  \REQUIRE in
%  \ENSURE  out
With $N_{\text{d}}$ IAB donors, $N_\text{c}$ IAB child nodes inside the network area, do the followings: 
\begin{enumerate}[I.]
\item Place the 1st node, $i = 1$, randomly in the considered network area.
\item Place the next node $i+1$ where $i = 1, 2, 3, ...., (N_\text{c} + N_\text{d}-1)$.
\item  Find the minimum inter-node distances $s_i$ between $(i+1)$th node and each of other nodes. 
\item If any $s_i < r_{\text{th}}$, redistribute the last node $(i+1)$th by repeating Steps II-IV until $s_i > r_{\text{th}}$.

\item For the obtained node locations, calculated the coverage. Then, proceed to Step I and continue the process for $N_{\text{it}}$ iterations pre-considered by the network designer, saving the best set of node locations $L_b$  among the considered solutions $L_j,\forall j, j = 1, 2, 3,..., N_{\text{it}},$ which gives in the best value of the service coverage. 
\end{enumerate}
\textit{Return the set of the node locations in Step V as the optimal node location set.}
\label{algorithm1}
\end{algorithm}

\begin{algorithm}[tbph]
\caption{IAB placement in the presence of constrained areas}
%  \REQUIRE in
%  \ENSURE  out
With $N_{\text{d}}$ IAB donors, $N_\text{c}$ IAB child nodes and a set of constrained areas inside the network area, do the followings: 
\begin{enumerate}[I.]
\item  Place the IAB donors/IAB nodes randomly in the considered network area.
\item  Identify the IAB node(s) falling inside the constrained areas. 
\item  For each of the nodes identified in Step II, redistribute the nodes. 
\item Proceed to Step II and continue the process until all \textcolor{black}{IAB} nodes fall outside the constrained areas. Save the set of locations as $L_i$.
For each selected possible node locations $L_i$, \textcolor{black}{compute the utility function} $\text{CP}_i$, $i=1,...., N_{\text{it}}.$
For instance, considering the service coverage, the objective function is given by \eqref{eq:rho}.

\item Proceed to Step I and continue the process for $N_{\text{it}}$ iterations pre-considered by the network designer, saving the best set of node locations $L_b$  among the considered solutions $L_i,\forall i,$ which gives in the best value of the utility function, e.g., service coverage.
\end{enumerate}
\textit{Return the set of the node locations in V as the optimal node location set.}

\label{algorithm2}
\end{algorithm}

In Algorithms 1 and 2, we propose greedy-based methods for IAB placement with minimum inter-IAB distance and geographical constraints, respectively. The algorithms are based on rejection-sampling method where multiple possible solutions are checked such that, satisfying the constraints, the service coverage is maximized.

Note that we present the algorithms for the general case where the position of both the IAB donors and the IAB nodes are optimized. However, in practice, the position of the IAB donor may be pre-determined based on, e.g., the fiber availability. Moreover, we present the algorithms for the simplest cases where each of the $N_\text{it}$ possible set of locations is determined independently. However, one can use, e.g., genetic algorithms to generate the new set of possible solutions based on, e.g., mutation of the previously obtained solutions \cite{9548327,adare2021uplink}. Such more complex algorithms may also be of interest in the cases with a large number of nodes. Finally, we present the setup for the cases finding a given number of possible solutions $N_\text{it}.$ Alternatively, one can run the algorithms until no further improvement is observed in a window of the obtained solutions. 

\section{Simulation Results and Discussion}
\label{res_sec}

In this section, we evaluate the effect of inter-node distance and the effect of constrained deployment optimization on the service coverage \eqref{eq:rho} of the IAB networks. 

% \begin{figure}[h!]

% \begin{subfigure}{\textwidth}
% \captionsetup{singlelinecheck = false, format= hang, justification=raggedright, labelsep=space} 
%     \includegraphics[width=3.5in\textwidth]{Figures/illustration_fig1.png}
%     \caption{subplot a.}
%     \label{fig:second}
% \end{subfigure}
% \begin{subfigure}{\textwidth}
% \captionsetup{singlelinecheck = false, format= hang, justification=raggedright, labelsep=space}    
%     \includegraphics[width=3.5in\textwidth]{Figures/Plot1_Line.png}
    
%     \caption{subplot b.}
%     \label{fig:third}
% \end{subfigure}
        
% \caption{Service coverage probability as a function of the distance from IAB donor to child IAB $s$ in subplot a with blockage $\lambda_{\text{bl}}$ = 500 $\text{km}^{-2}$.}
% \label{linebs1}
% \end{figure}
\begin{figure}
\begin{minipage}{\columnwidth}
\begin{center}
\includegraphics[width=4in]{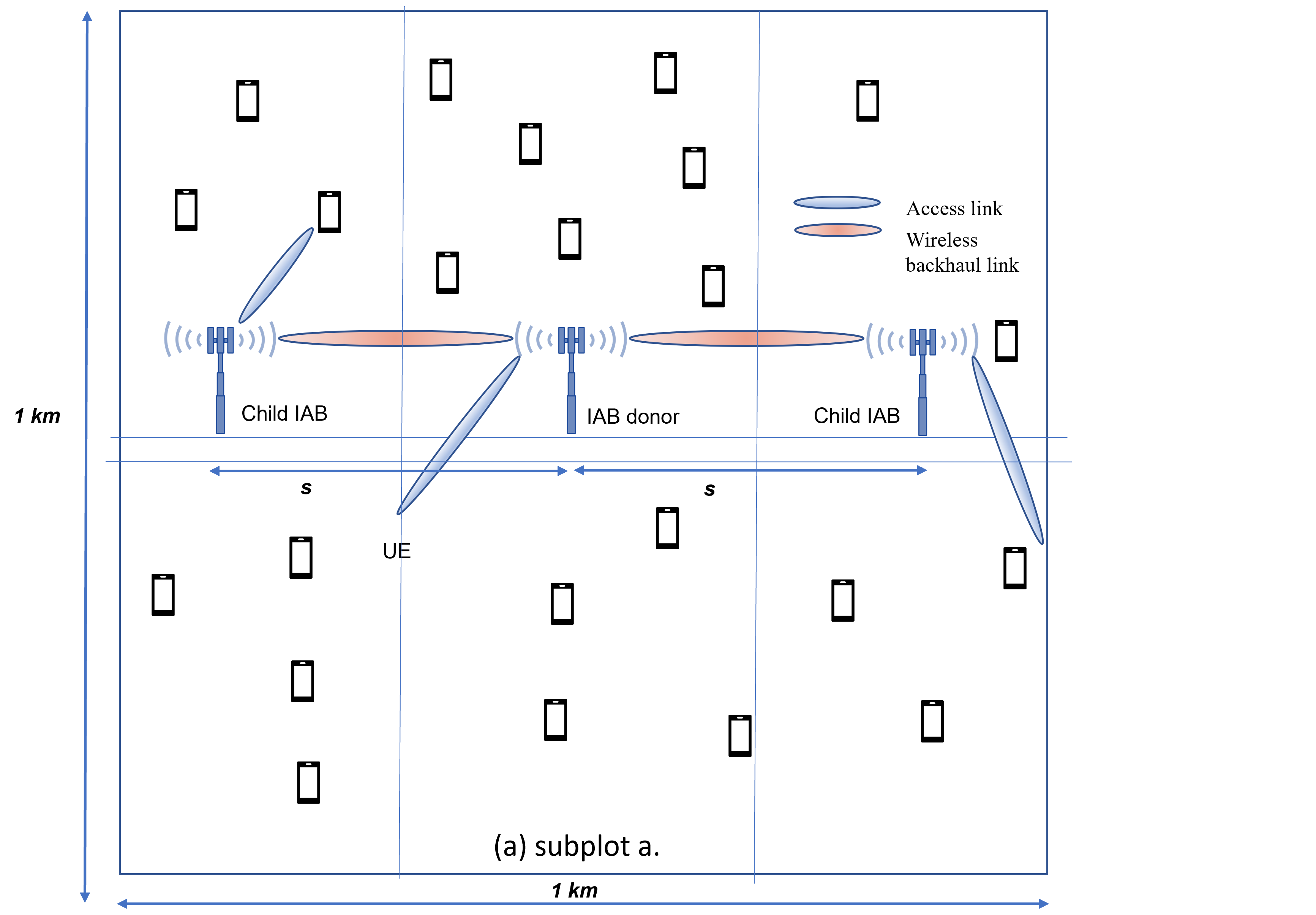}
\end{center}

\end{minipage}
\begin{minipage}{\columnwidth}
\includegraphics[width=3.5in]{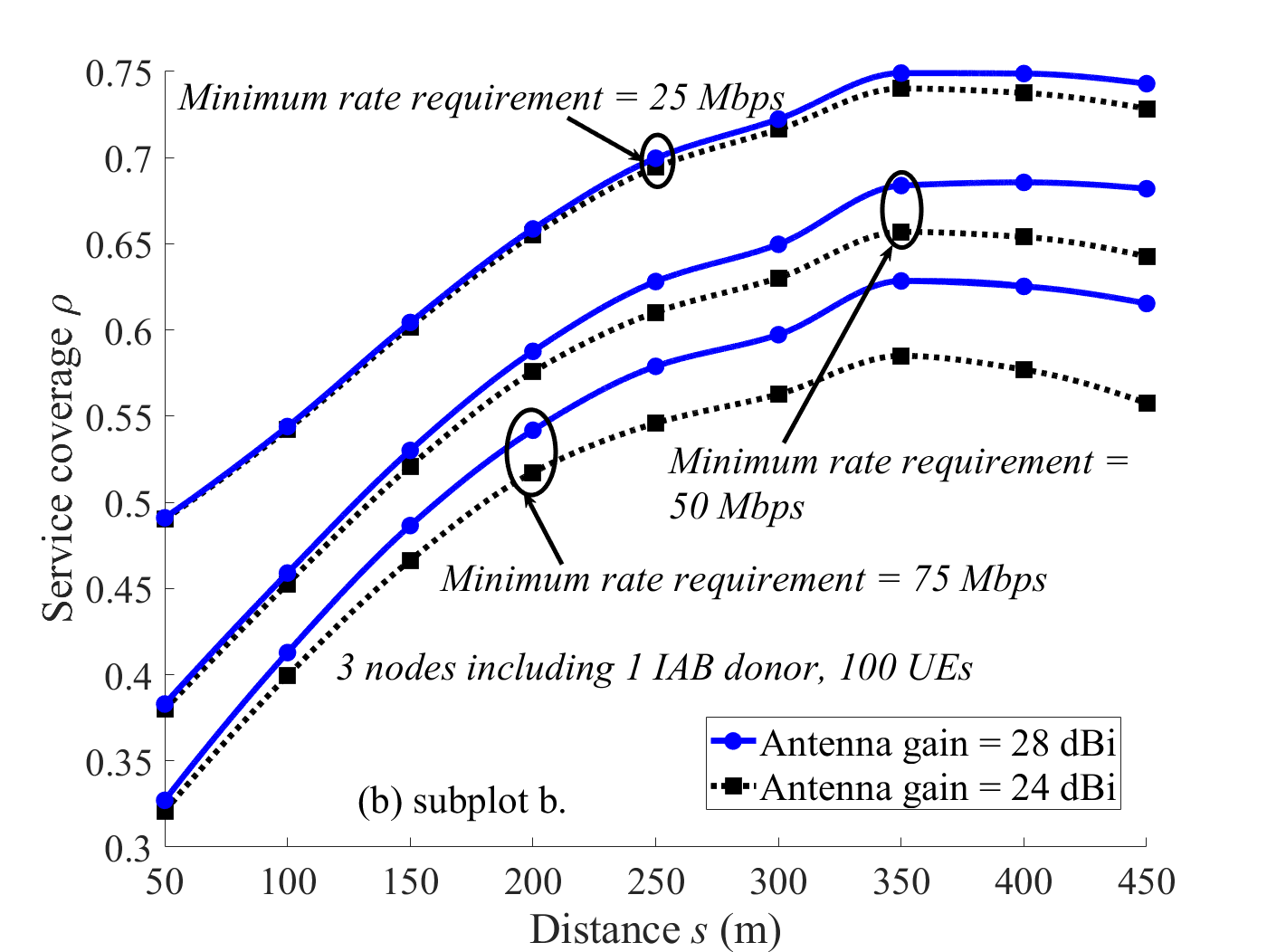}
\end{minipage}
\caption{Service coverage as a function of the distance from IAB donor to child IAB $s$ in subplot a with blockage $\lambda_{\text{bl}}$ = 500 $\text{km}^{-2}$.}
\label{linebs1}
\end{figure}
\begin{figure}
\centerline{\includegraphics[width=3.5in]{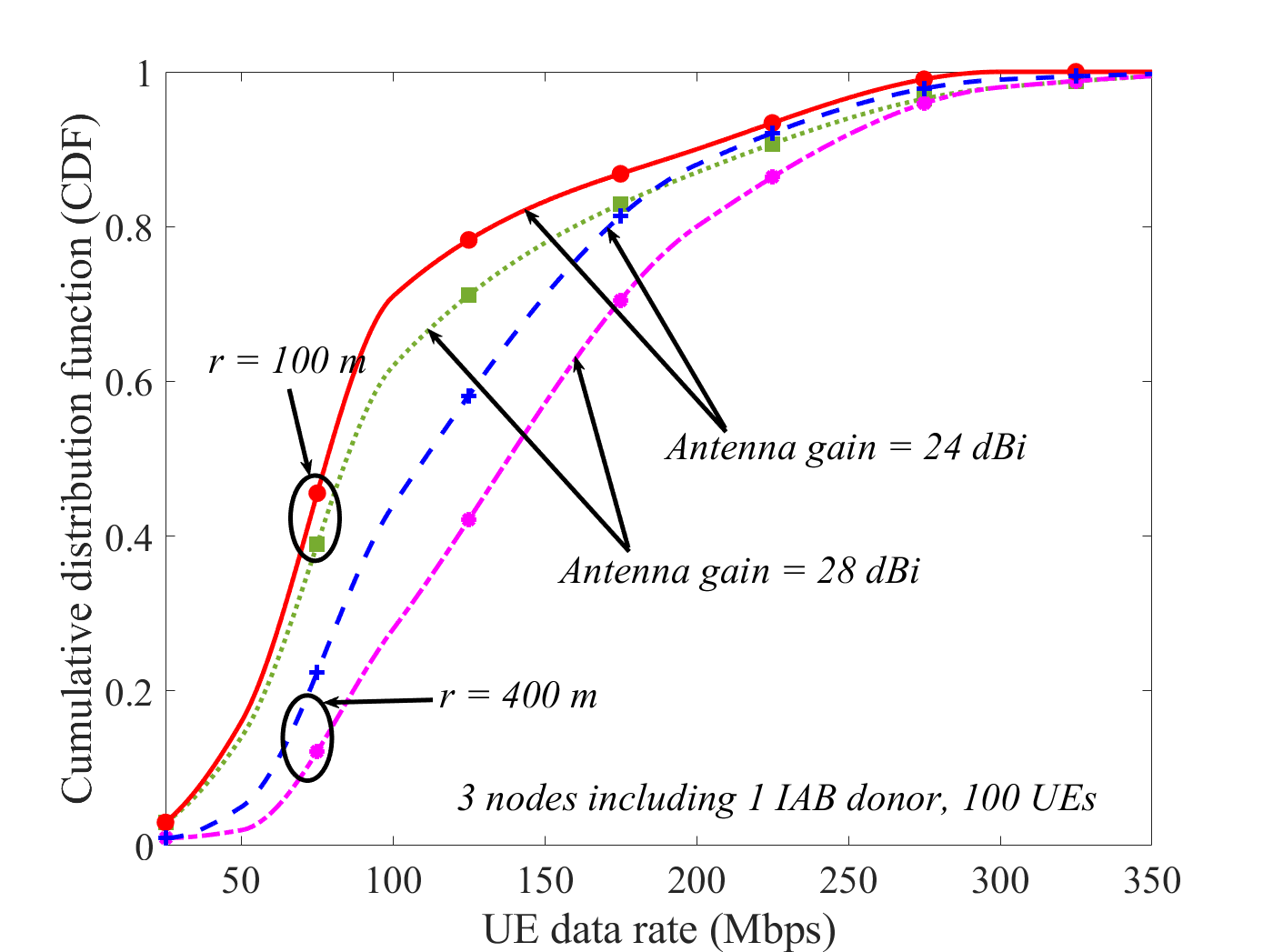}}
\caption{CDF of the achievable rates with blockage density $\lambda_{\text{bl}}$ = 500 $\text{km}^{-2}$, and 100 UEs.}
\label{circlebs}
\end{figure}
% \begin{center}
% \includegraphics{yourimage}
% \end{center}

% \begin{figure}
% \centerline{\includegraphics[width=3.5in]{Figures/minLine_ReducedUEdist.png}}
% \caption{\textcolor{black}{Service coverage probability as a function of the distance from center BS to BSs on side with $\lambda_{\text{B}}$ = 500 $\text{km}^{−2}.$}}
% \label{linebs}
% \end{figure}
% \begin{figure}[h!]

% \begin{subfigure}{\textwidth}
% \captionsetup{singlelinecheck = false, format= hang, justification=raggedright} 
%     \includegraphics[width=3.3in\textwidth]{Figures/Fig_circlesystem1.png}
%     \caption{subplot a.}
%     \label{fig:second}
% \end{subfigure}
% \begin{subfigure}{\textwidth}
% \captionsetup{singlelinecheck = false, format= hang, justification=raggedright}    
%     \includegraphics[width=3.5in\textwidth]{Figures/minCirclev2.png}
    
%     \caption{subplot b.}
%     \label{fig:third}
% \end{subfigure}
% \caption{Service coverage probability as a function of the minimum distance constraint between the nodes, blockage density $\lambda_{\text{bl}}$ = 500 $\text{km}^{-2}$, child IAB node density $\lambda_{\text{child}}$ = 20 $\text{km}^{-2}$.}
% \label{linebs4}
% \end{figure}
\begin{figure}
\begin{minipage}{\columnwidth}
\includegraphics[width=3.2in]{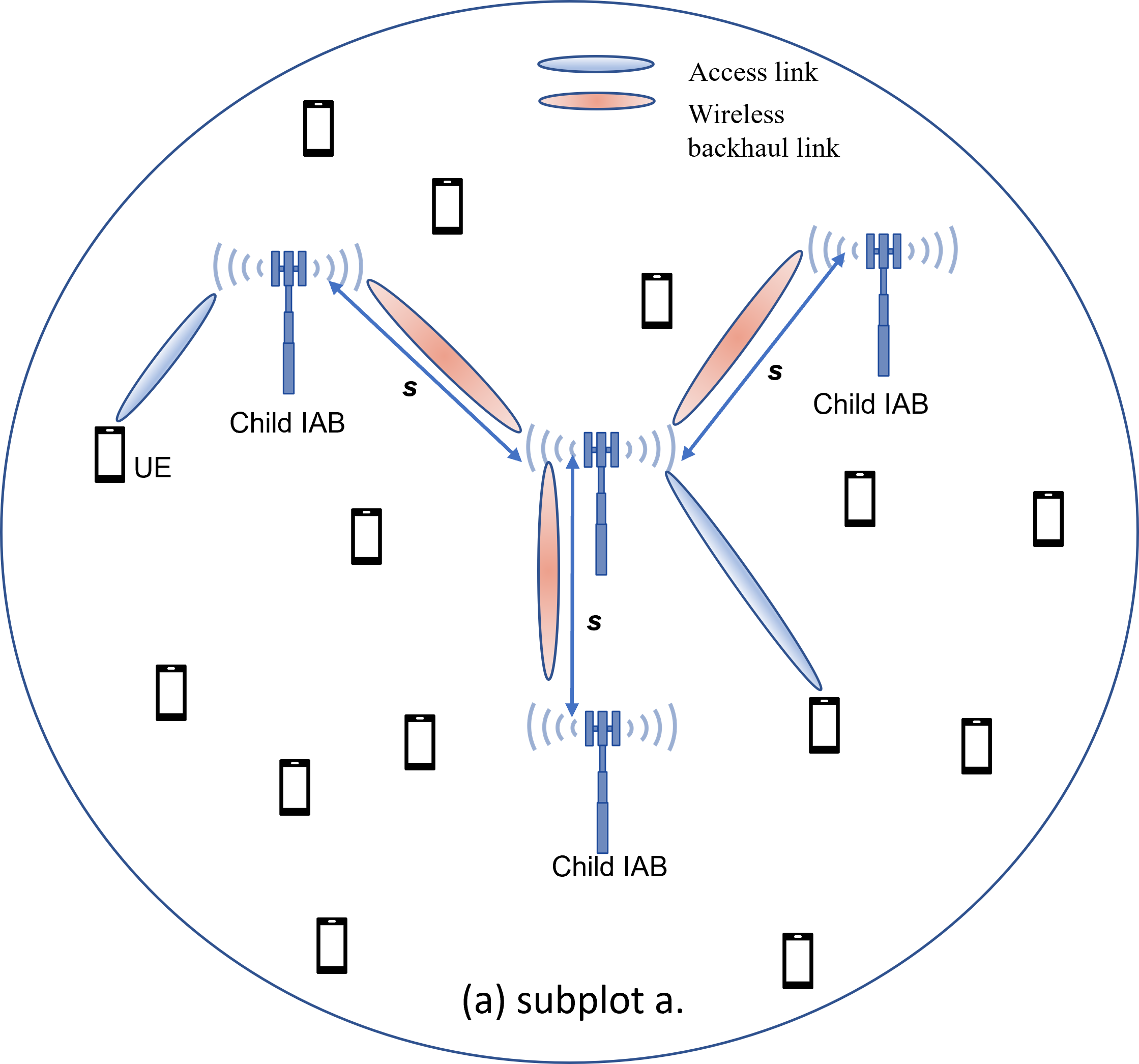}
\end{minipage}
\begin{minipage}{\columnwidth}
\includegraphics[width=3.5in]{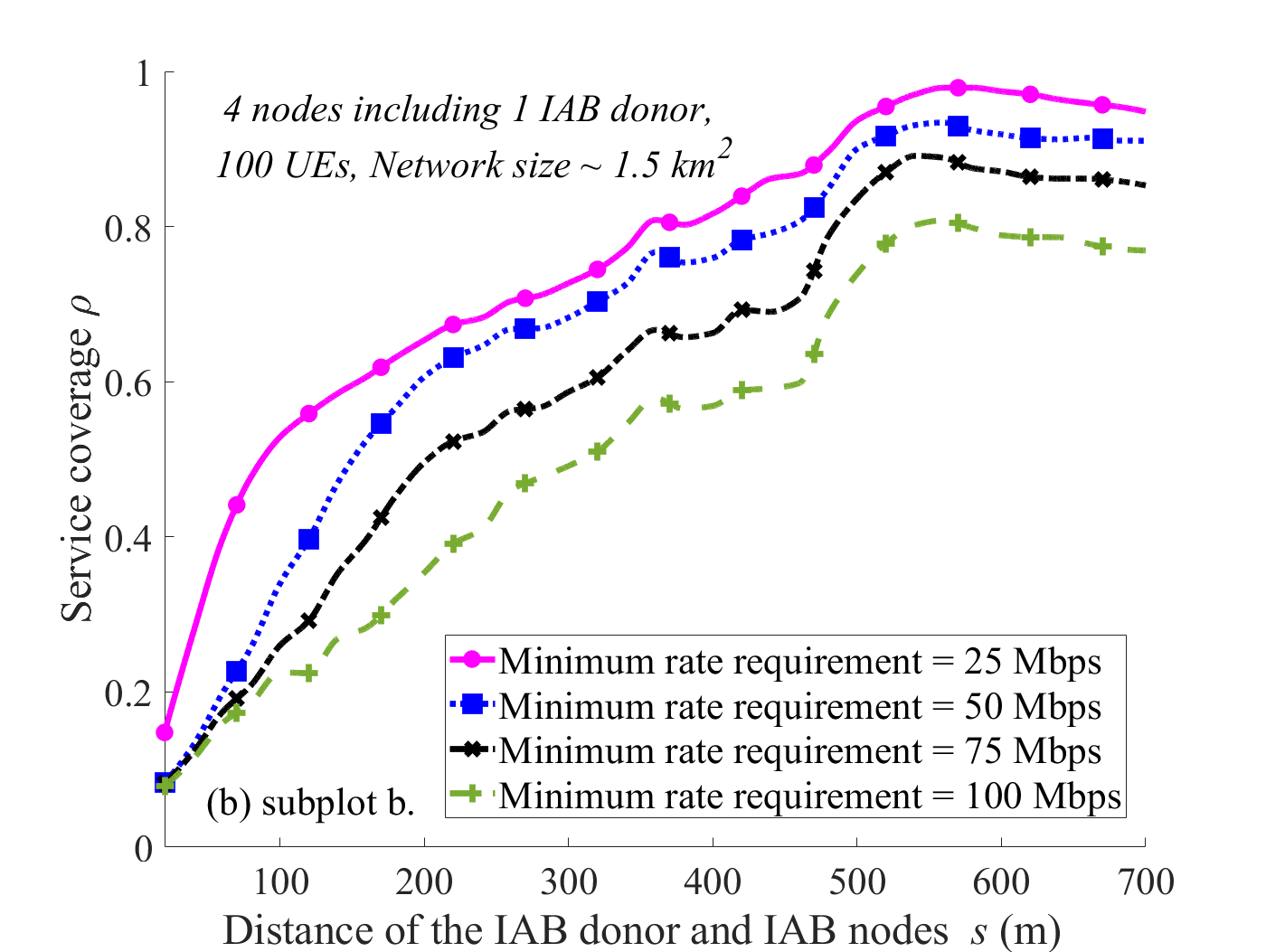}
\end{minipage}
\caption{Service coverage as a function of the minimum distance constraint between the nodes, blockage density $\lambda_{\text{bl}}$ = 500 $\text{km}^{-2}$, child IAB node density $\lambda_{\text{child}}$ = 20 $\text{km}^{-2}$.}
\label{linebs4}
\end{figure}
\begin{figure}
\centerline{\includegraphics[width=3.5in]{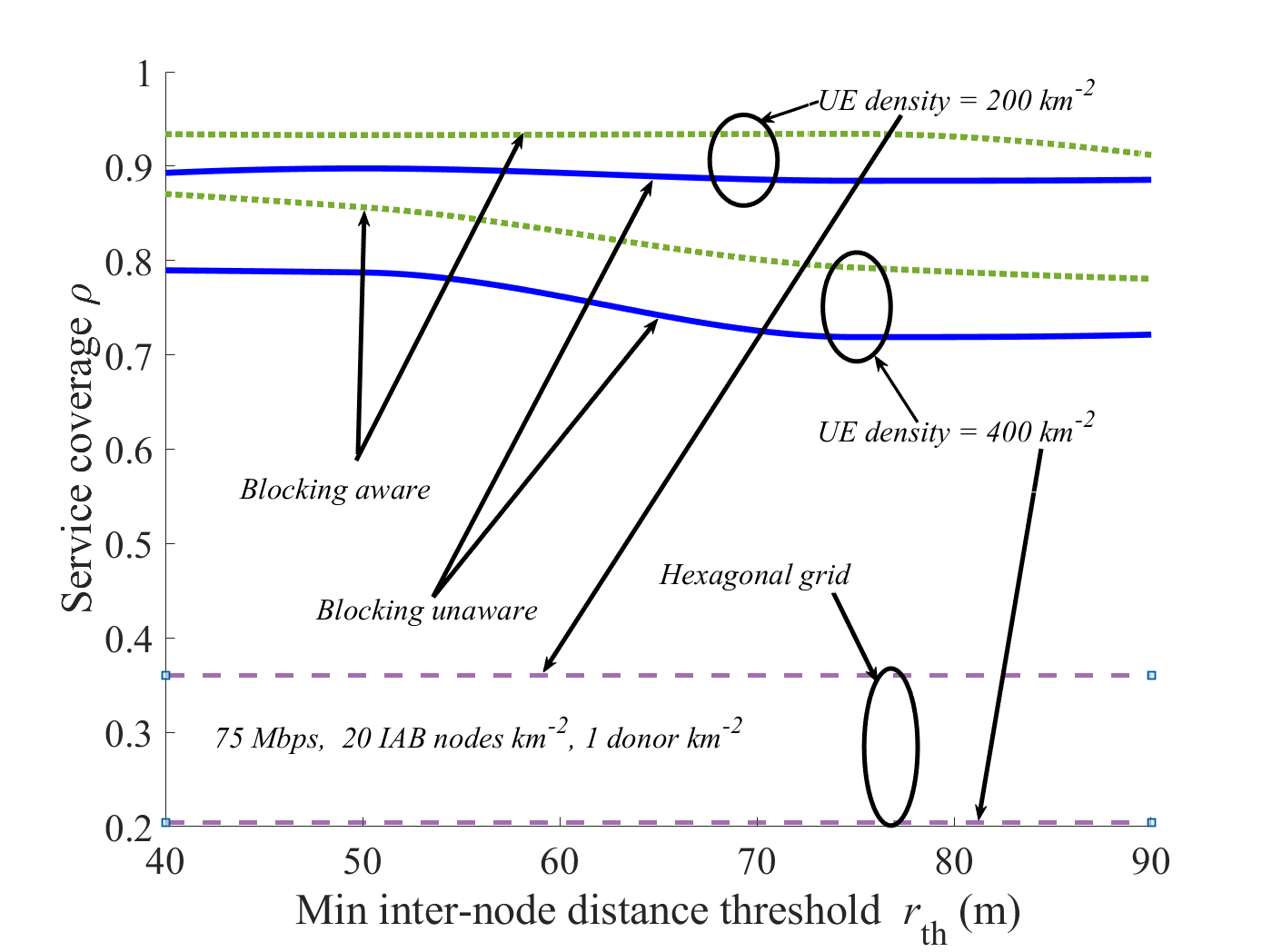}}
\caption{Service coverage as a function of the minimum distance constraint between the nodes, blockage density $\lambda_{\text{bl}}$ = 500 $\text{km}^{-2}$, child IAB node density $\lambda_{\text{child}}$ = 20 $\text{km}^{-2}$.}
\label{minradiusfig}
\end{figure}
%\begin{figure}
%\centerline{\includegraphics[width=3.5in]{Figures/system_hexagonal.png}}
%\caption{Nodes in hexagonal grid.}
%\label{circle3hexa}
%\end{figure}

% \begin{figure}
% \centerline{\includegraphics[width=3.5in]{Figures/minCirclev2.png}}
% \caption{\textcolor{black}{Service coverage probability as a function of the distance of BSs from center fiber-connected BS with $\lambda_{\text{B}}$ = 500 $\text{km}^{−2}.$}}
% \label{circlebs}
% \end{figure}
Figure \ref{linebs1}b demonstrates the service coverage as a function of the distance between the IAB donor and child IAB nodes, $s$ as of the symmetric setup shown in Fig. \ref{linebs1}a of which the donor is located at the center and child IAB nodes are placed symmetrically besides. As shown, the service coverage improves as the IAB nodes are well distributed in the area up to certain distance. Intuitively, this is supported by the decreased interference among the nodes and also better coverage in the area. However, the coverage later starts to drop at large values of $s,$ due to the low coverage experienced by the UEs in the middle of the IAB donor and child IAB nodes.

In Fig. \ref{circlebs}, we study the cumulative distribution
function (CDF) of the UEs achievable data rates in the cases with different antenna gains and inter-node distances, $s = 100$ m and $s = 400 $ m. Here, the parameters are set to UE density = 100 $\text{km}^{-2}$, $P_m, P_s$ = {24 dBm}, and the IAB-donor is located at the center. As can be seen in Fig. \ref{circlebs},  higher antenna gain gives the opportunity to support higher access data rates depending on the inter-node distances. For instance, with $G_m, G_s$ = 28 dBi and $s = 400$ m around 20\% of UEs may experience $> 200$ Mbps access rates, compared to the 13\%, when  $G_m, G_s$ = 24 dBi. Moreover, the effect of the antennas gain increases with the inter-node distance (Figs. \ref{linebs1} and \ref{circlebs}).

Figure \ref{linebs4}b demonstrates the service coverage as a function of the distance between the IAB donor and child IAB nodes, $s$ as of the symmetric setup shown in Fig. \ref{linebs4}a of which the IAB donor is located at the center and child IAB nodes are placed symmetrically with equal distance from the donor. As shown, the service coverage increases with the node separation $s$, up to a point around 550 m, which is due to the decreased interference between the nodes and at the same time properly covering the area. Then, the coverage starts to slightly drop due to the coverage reduction for the UEs in between too far nodes. In this way, there is an optimal distance between the nodes maximizing the coverage. Finally, the coverage decreases significantly with increased UEs minimum rate requirements, to compensate of which one needs more resources/IAB nodes.

In Fig. \ref{minradiusfig}, we study the effect of deployment optimization. Particularly, considering a minimum inter-node distance constraint, we compare the coverage of the IAB networks in the cases with optimized deployment, optimized by Algorithm 1, and the cases with hexagonal IAB deployment.

Here, the results are presented for the cases where the IAB donor has $G_m = 24$ dBi and child IAB nodes have a gain of $G_s$ = 18 dBi for IAB nodes density of 20 $\text{km}^{-2}$. Moreover, the figure presents the results for the cases where the nodes locations are obtained only by considering the minimum distance between them or when the blockages and the backhaul links' qualities are also taken into account in the optimization. As we see, the service coverage drops when the constraint becomes tighter, however, for all considered range of constraints, compared to hexagonal deployment, constrained deployment optimization increases the network coverage significantly. Indeed, knowing the blockages locations helps in improving the deployment optimization, specially when the UE density increases. Finally, the effect of inter-node distance constraint on the coverage increases with the UE density.

\begin{figure}
\centerline{\includegraphics[width=3.5in]{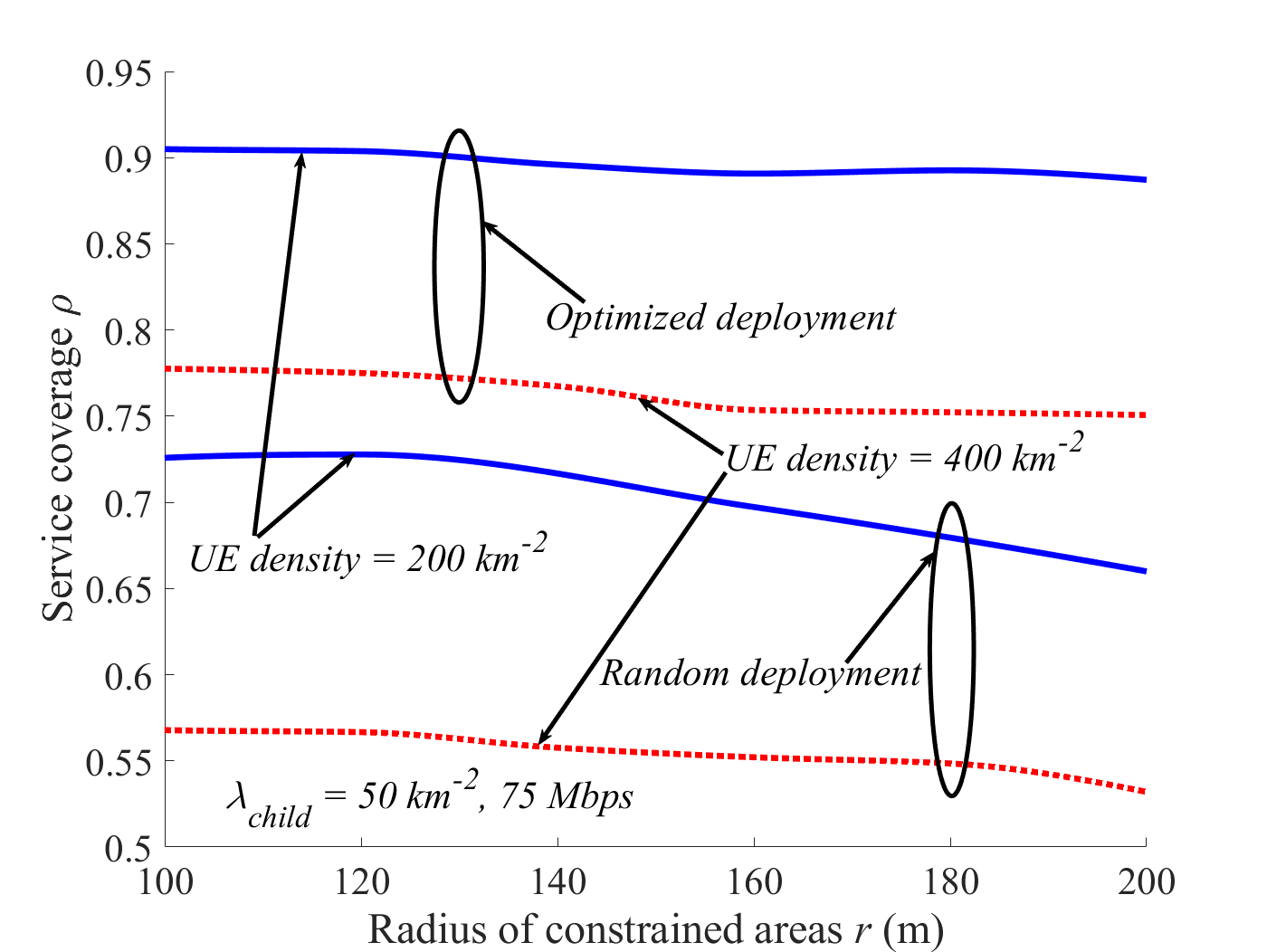}}
\caption{Service coverage as a function of the radius of the constrained areas ($c$), blockage density $\lambda_{\text{bl}}$ = 500 $\text{km}^{-2}$, $\lambda_{\text{child}}$ = 50 $\text{km}^{-2}$.}
\label{circle2}
\end{figure}

Figure \ref{circle2} verifies the effect of geographical constraints, on the coverage of IAB networks. Particularly, we study the coverage of the deployment-optimized IAB networks in the cases where, following Fig. 1b, the IAB nodes can not be placed in constrained areas. Here, the results are presented for a network consisting of five circular constrained areas of radius $c$, with blockage density $\lambda_\text{bl} = 500 \text{ km}^{-2}$, child IAB node density $\lambda_\text{child} = 50 \text{km}^{-2}$, and minimum rate requirement $R_U = 75$ Mbps. The results are presented for the radius of each constrained areas ranging from 100 m to 200 m which corresponds from  10\% to 40\% of the total disk area, respectively.

As demonstrated in Fig. \ref{circle2}, with low geographical constraints, network performance is not affected by the deployment constraints. However, with large area constraints, the service coverage decreases. This is intuitively because, with larger constraints for IAB placement, there is an increased chance of low coverage for users within the constrained areas. Also, since the IAB nodes get packed outside the constrained areas, interference levels for the UEs outside the constrained areas increases resulting in a further decrease in coverage. It can be seen in Fig. \ref{circle2}, where the optimized IAB network in the presence of  UE density = 200 $\text{km}^{-2}$ increases the coverage to 90.5\% from the case with UE density = 400 $\text{km}^{-2}$ with coverage of 77\%. Finally, compared to random deployment, proper network planning boosts the coverage significantly. Also, compared to the case with child IAB nodes distributed randomly in the unconstrained areas, the coverage is less severely affected by geographical constraints when optimized by Algorithm \ref{algorithm2}.

\section{Conclusion}
We studied the problem of IAB network deployment optimization in the cases with different deployment constraints. We proposed iterative constrained deployment optimization methods with no need for mathematical analysis and with the capability to be adapted for different channel models/constraints/metrics of interest. As demonstrated, with different geographical and inter-node distance constraints, compared to random or hexagonal deployments, proper network planning can boost the coverage of the IAB networks significantly.  Finally, in practice, deployment planning may be affected by, e.g., the availability of non-IAB backhaul connection in specific areas, local authority regulations, and the designer
may consider, e.g., the planned infrastructure changes, cost, seasonal variations.

\section*{Acknowledgment}

This work was supported in part by the European Commission through the H2020 Project Hexa-X under Grant 101015956, and in part by the Gigahertz-ChaseOn Bridge Center at Chalmers in a project financed by Chalmers, Ericsson, and Qamcom.

\bibliographystyle{IEEEtran}
\bibliography{bibliography}

\end{document}